\documentclass{article}
\usepackage{eurosym}
\usepackage{amsmath,amssymb,amsthm}
\usepackage{graphicx}
\usepackage{subcaption}
\usepackage{verbatim}
\usepackage{bm}
\usepackage{color}
\usepackage{amsmath}
\usepackage{amsfonts}
\usepackage{amssymb}

\begin{document}

\title{An elastically stabilized spherical invagination}
\author{Xiaoyu Zheng$^{1}$\thanks{%
Email: xzheng3@kent.edu}, Tianyi Guo$^{2}$ \thanks{%
Email: tguo2@kent.edu}, Peter Palffy-Muhoray$^{1,2}$\thanks{%
Corresponding author Email: mpalffy@kent.edu} \\
%EndAName
\emph{$^1$Department of Mathematical Sciences, Kent State University, OH,
USA }\\
\emph{$^2$Advanced Materials and Liquid Crystal Institute, Kent State
University, OH, USA} }
\maketitle

\begin{abstract}
    Invaginations are partial enclosures formed by surfaces. Typically formed by biological membranes; they abound in nature.  
In this paper, we consider fundamentally different structures: elastically stabilized invaginations. Focusing on spherical 
invaginations formed by elastic membranes, we carried out experiments and mathematical modeling to understand the 
stress and strain fields underlying stable structures. Friction plays a key role in stabilization, and consequently the required
force balance is an inequality.  Using a novel scheme, we were able to find stable solutions of the balance equations for 
different models of elasticity, with reasonable agreement with experiments.
\end{abstract}


\section{Introduction}

\begin{figure}[htb]
	\centering
	\includegraphics[width=.4\linewidth]{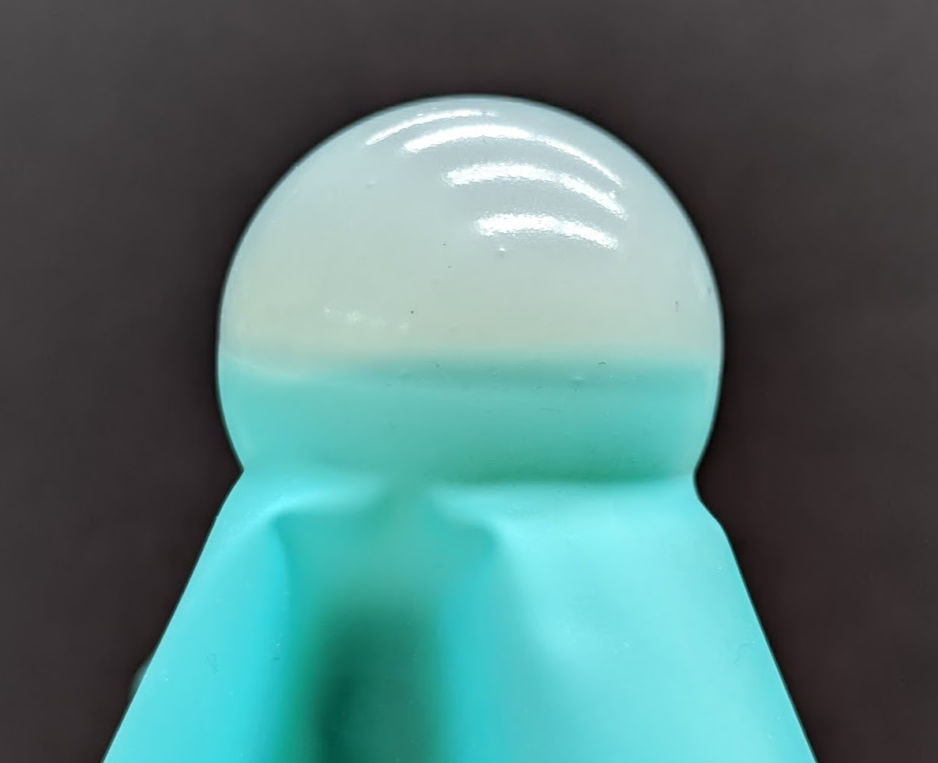}  
	\caption{An elastically stabilized spherical invagination formed by a dental
		dam and a table tennis ball.}
	\label{fig:sphex}
\end{figure}

Invaginations are cavities with narrow openings; they are partial enclosures
formed by surfaces. They abound in nature. In the process of pinocytosis 
\cite{Stillwell}, a cell absorbs particles - nutrients - by surrounding them
with its membrane, first forming an invagination partially enclosing the
particle and subsequently completely enclosing and engulfing it. Such
invaginations are typically created by membranes and not by the passive
partially enclosed object. Here we consider a different process, where
the deformation of an elastic membrane is the result of forces exerted by a
solid body resulting in an invagination entrapping the deforming
body. An elastically stabilized deformation of an elastic membrane
surrounding a rigid disk has already been considered \cite{Magic}.
Motivation for this work is to explore the effects of geometry. Here we
study a spherical invagination in an elastic membrane surrounding a rigid
sphere, as shown in Fig.~\ref{fig:sphex}, both experimentally and theoretically.
We first describe our experiments in producing and characterizing spherical
invaginations. We next discuss simple mathematical models and numerical
simulations. We close by summarizing our findings.

\section{Experiments}

\subsection{The membrane and the sphere}

The elastic membranes used in our experiments were thin natural latex sheets
sold as powder-free Crosstex 19202 dental dams. Their dimensions are $%
5^{\prime \prime }\times 5^{\prime \prime }\times 150\mu m$. Their elastic
modulus is $2MPa\footnote{
The modulus and thickness of the membranes were measured by us.}$. The
maximum stretch we were able to achieve in one direction without damaging
the membrane is $6$. The spheres used in the experiments discussed were
regulation size $40mm$ dia.~table tennis balls. Since friction between the
ball and the membrane plays a key role in stabilizing the invagination,
we attempted to measure both the static $\mu _{s}$ and the kinetic $\mu _{k}$ coefficients of
friction between the membrane and the ball. For
the measurements, we used narrow strips ($15cm\times 5mm$) of the membrane.
For the static measurements, we attached a mass at each end of the strip
supported by the ball. We increased the mass at one end until the membrane
started slip. The friction coefficient $\mu _{s}$ can be obtained from the ratio of the masses via
the Euler capstan equation \cite{Euler}, 
\begin{equation}  \label{Euler}
T_{load}=T_{hold}\exp(\mu\theta),
\end{equation}
where $T$ is the tension, $\theta$ is the subtended angle.
For the kinetic measurements, we used a lathe to rotate the table tennis
ball while it supported the strip. One end of the strip was fixed and held horizontal while a mass was suspended from
the other. We measured the strain of the strip near the fixed end while ball
was rotating, $\mu _{k}$ can be obtained from the ratio of the force
producing the strain and the weight of the mass again via Eq.~\eqref{Euler}.
Our measurements gave $\mu _{s}=0.99$ and $\mu _{k}=0.82$. We were unable to
find results for the kinetic coefficient in the literature, but our value
for the static coefficient is in the range $0.86<\mu_k <2.18$ reported in \cite%
{static}.

\subsection{Forming the invagination}

The invagination by the membrane was realized using two distinct methods.

1. Our original method consisted of stretching the central portion of the
membrane in plane, by hand, essentially producing an extremely highly
stretched horizontal circular region, approximately $5^{\prime \prime }\times
5^{\prime \prime }$, in the center of the original sheet. This was then
placed above the table tennis ball, which was supported by the neck of a
glass bottle with $1^{\prime \prime }$ OD opening. The stretched membrane
was lowered onto the ball, maintaining cylindrical symmetry and maintaining the radial stretch as the edges were
further lowered below and beneath the ball; necking of the stretched
membrane below the ball could be clearly  observed. The tension was then
slowly released, with the membrane contracting and slipping on the ball,
coming to rest and forming the invagination shown in Fig. \ref{fig:sphex}.

\begin{figure}[htb]
	\centering
	(a) \includegraphics[width=.3\linewidth]{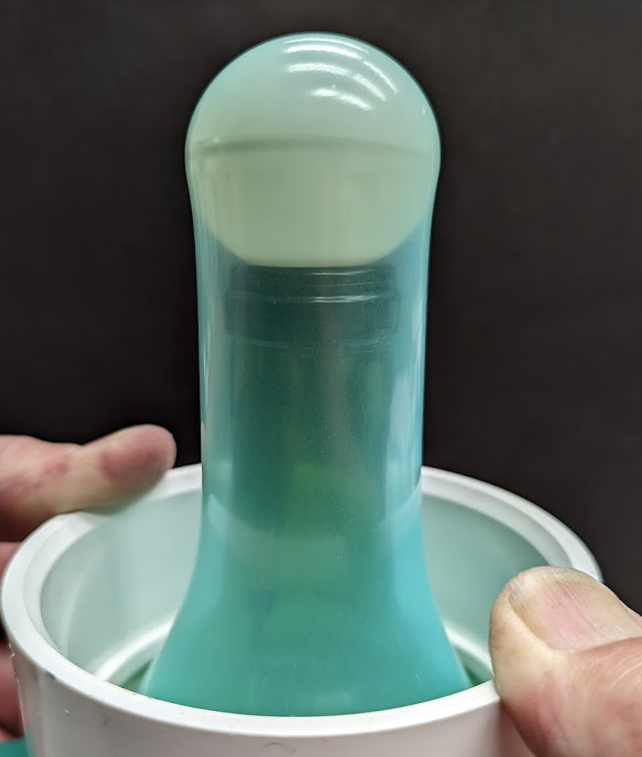}  \qquad\qquad
	(b) \includegraphics[width=.3\linewidth]{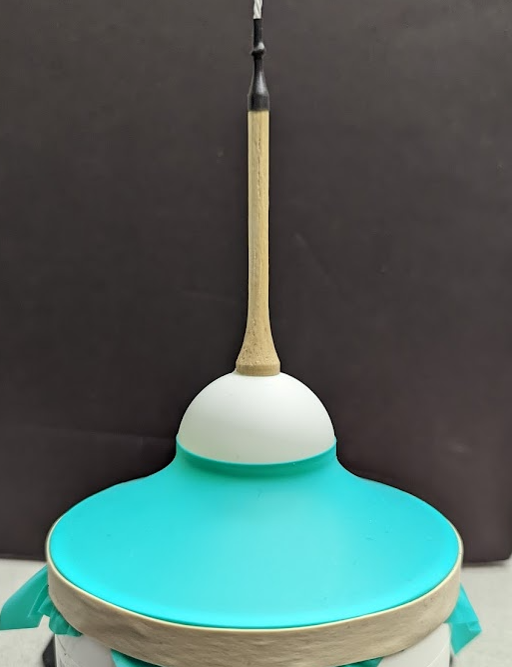}  
	\caption{(a) Forming the invagination by pushing down the PVC tube with membrane held in place at the bottom. (b) Removing the sphere from the invagination by pulling the rod attached to the ball.}
	\label{fig:tee}
\end{figure}

2. Here the unstretched membrane was placed on one end of a $4^{\prime
\prime }$ high $4^{\prime \prime }$ OD PVC tube. The membrane was held
in place by a $1/2^{\prime \prime }$ wide rubber band pressing the edges
of the membrane to the tube. With the tube vertical and the membrane at the
bottom end, the membrane again was placed above the table tennis ball,
supported as before. The center of the ball was on axis of the tube. As the
tube was lowered, the membrane stretched and slipped on the ball, 
eventually becoming highly stretched, as shown in Fig.~\ref{fig:tee}(a). The tube was then slowly raised, with
the membrane contracting and slipping on the ball, coming to rest and
forming the invagination shown in Fig.~\ref{fig:sphex}.

\begin{figure}[htb]
	\centering
	(a)\includegraphics[width=0.4\linewidth]{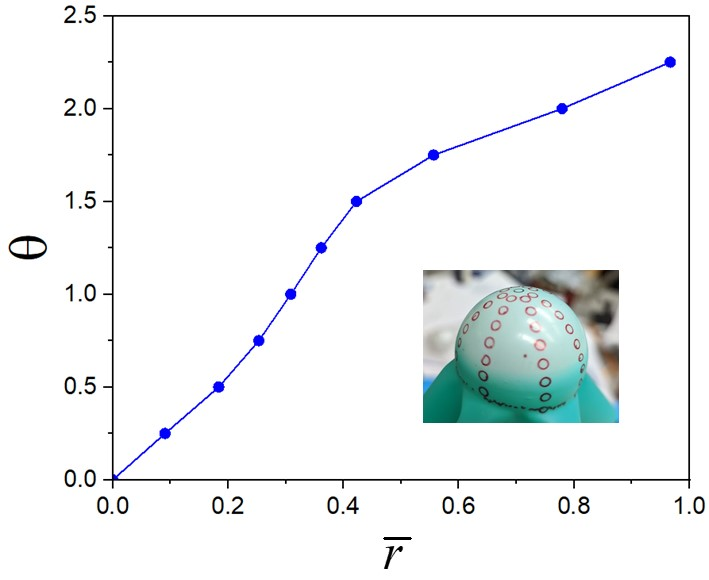} \qquad
	(b)\includegraphics[width=0.4\linewidth]{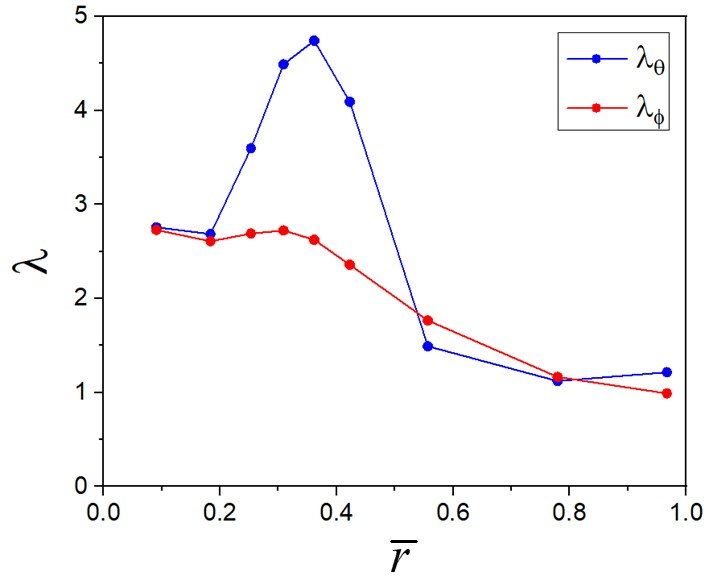}
	\caption{Experimental results. (a) Deformation $\theta$ vs.~$\bar{r}$. The inset shows the circles inscribed on the deformed membrane.  (b) Principal stretches $\lambda_{\theta}$ and $\lambda_{\phi}$ vs.~$\bar{r}$. Here both the deformation $\theta$ and undeformed distance $\bar{r}$ are normalized by the radius of the sphere $R$.
	}
	\label{fig:exp}
\end{figure}

\subsection{Releasing the sphere}

We were interested in determining the force required to release the sphere
from the invagination. We attached a solid rod (golf tee) to the ball held
by the membrane attached to PVC tube as shown in Fig.~\ref{fig:tee}(b), and via
pulleys and weights, applied an increasing upward force to the rod. In this
instance, it required a force of $2.3N$ to remove the table tennis ball from
the invagination.

\subsection{Characterizing the deformation}
To characterize the deformation of the membrane, using a template, we
inscribed circles, separated by $0.25$ radians along the meridional direction,
on the stretched membrane, as shown in the inset of Fig.~\ref{fig:exp}(a). The membrane
was removed from the ball, and the image on the undeformed membrane was
analyzed to characterize the deformation. The displacement $s$ of a point at distance 
$r$ from the center of the undeformed membrane was measured and both are normalized by the radius of the sphere, specifying the deformation $\theta (\bar{r})$, as shown in
Fig.~\ref{fig:exp}(a). Fig.~\ref%
{fig:exp}(b) shows the principal stretches, as defined by Eqs.~%
\eqref{eq_lambda_th} and \eqref{eq_lambda_phi}.

Although the two methods of realizing the invagination differ in that method
1 imposes a strong pre-stretch while method 2 does not, our measurements
indicate that the final configurations are nonetheless essentially the same.

\section{Modeling the invagination}

In this section, we present the mathematical models which describe the
deformation of the elastic membrane on the sphere involving Coulomb
friction. When the membrane loses contact with the sphere, it is still in
tension and is strictly under the sphere without making contact with it. The tension in the membrane rapidly relaxes to zero, below which point the
membrane is stress free. In this study, we focus only on the portion of the membrane which is in contact with the sphere. We first describe the force balance equation on
the elastic membrane, and then present the hyper-elastic models used to
calculate the forces, and finally we present our algorithm for the solution, followed by
numerical results.

\subsection{Force balance}

Our goal here is to identify the forces at play and to examine the
conditions for the stability of the invagination. The system under
consideration is a rigid sphere, partially covered by an elastic membrane as
shown in Fig.~\ref{fig:sphex}. The configuration there is stable; the
sphere is entrapped by the membrane and external forces must be applied to
separate the two.

\begin{figure}[htb]
\centering
\includegraphics[width=.4\linewidth]{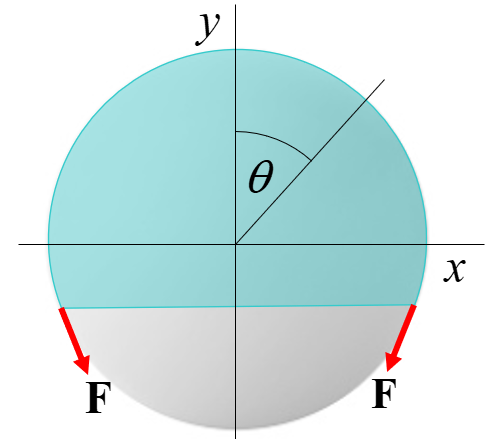}
\caption{Schematic of the invagination. The force shown is exerted while the membrane is being stretched. Only the portion of the membrane which is in contact with the sphere is shown. }
\label{fig:schematic}
\end{figure}

A schematic is shown in Fig.~\ref{fig:schematic}. The sphere with 
radius $R$ is at rest with its center at the origin. Its surface exerts not only
a normal force, but also a tangential friction force, with friction
coefficient $\mu $. The initially undeformed elastic membrane is flat,
regarded here for simplicity as circular in shape, with thickness $h_{0}$. It is then stretched and placed on the sphere so as to
cover a portion of the sphere, as shown in Fig.~\ref{fig:schematic}.  The experimental details for
producing the stable invagination are given in Section 2.

We assume that the system has cylindrical symmetry about the $y$-axis. We
denote the points on top and bottom of the sphere as the north and south
poles, respectively; great circles passing through the poles form the meridians along
which latitude is measured, perpendicular to these are the lines of
latitude, along which the azimuth is measured. Using the standard spherical
coordinate notation, distance $s$ from the north pole is $R$ multiplied by the polar
angle $\theta$ along a meridian, and points along the azimuth are
identified by the angle $\phi$. The undeformed elastic membrane is a flat
disc, where distance from the center in the plane is measured by $r$,
distance perpendicular to the plane is measured by $z$.

When the deformed membrane forming the invagination is on the sphere and the
system is in equilibrium, the net force on any element of the membrane must
vanish. There will be stress $\boldsymbol{\sigma }$ in the membrane, with
principal stress vectors in the tangential meridional, the tangential
azimuthal and the normal directions. The net force on any element of the
membrane with volume $\Delta V$ exerted by the surrounding membrane is 
\begin{equation}
\mathbf{F}_{sm}=-\int_{\Delta V}\nabla \cdot \boldsymbol{\sigma }dV,
\end{equation}%
with components in all three principal directions. In addition to $\mathbf{F}%
_{sm}$, there will be a normal force $\mathbf{F}_{N}$ per area exerted by the
sphere on the membrane, normal to the sphere, and a friction force $\mathbf{F%
}_{f}$ per area in the meridional direction in equilibrium.  Due to
cylindrical symmetry, there will be no friction force in the azimuthal
direction. If the stress vectors in the meridional and azimuthal directions
are multiplied by the thickness of the
membrane divided by the sphere radius, they give $T_{\theta }$ and $T_{\phi }$, the components of the
tangential force per area due to stress on the membrane in the meridional
and azimuthal directions, respectively.

The net force in the azimuthal direction is zero due to symmetry and the net
force in the normal direction is zero by Newton's third law. Force balance
in the meridional direction requires that%
\begin{equation}\label{FB}
F_m=-{F}_{f},  
\end{equation}%
where ${F}_{f}$ denotes
the (areal) friction force density, and
\begin{equation}
F_{m}=\frac{\partial T_{\theta }}{\partial \theta }+(T_{\theta }-T_{\phi
})\cot \theta 
\end{equation}
denotes the elastic force density along the meridian. 

Friction plays a key role in the system. During the formation of
the invagination, initially, as the membrane is pulled downward, it
slips on the sphere with $s(r)$ increasing in the northern hemisphere
for both methods 1 and 2.  After the stretching, as the tension in the
membrane is reduced, the membrane slips again - but now towards the north
pole with $s(r)$ decreasing in some - but not all - regions on the
sphere.  As the membrane slips, a tangential friction force density in the
meridional direction ${F}_{f}=\mu _{k}F_{N} $ is exerted by the sphere on
the membrane - the sign of the force depends on the direction of the slip;
the magnitude is determined, in the simplest approximation, by the kinetic friction coefficient. The membrane moves as long as the elastic force
density $F_{m}$ is greater than ${F}_{f}$. As the membrane slips, $F_{m}$
decreases, and when $F_{m}<$ ${F}_{f}$, the membrane comes to rest at that
point. To move the membrane at rest requires a force greater than the
maximum provided by static friction. 

Experimental observations indicate that the sliding membrane does not come
to rest all at once; regions near the equator continue slipping, while the
region near the north pole is at rest. Although the dynamics of the boundary
between the stationary and moving parts of the membrane is intriguing, due
to the complexity of dynamic friction \cite{Liu}, we will not consider it
here. For simplicity, we ignore the difference between static and kinetic
friction, assume that $\mu _{k}\approx \mu _{s}=\mu $, but stress that in
the static case, the relation ${F}_{f}=\mu F_{N}$ denotes only the maximum
possible friction force density that the sphere can provide. Explicitly, the
friction force density $F_{f}$ satisfies%
\begin{equation}
|F_{f}|\leq \mu F_{N},  \label{eq_inequality}
\end{equation}%
where, in our case, the normal force density is given by 
\begin{equation}
F_{N}=T_{\theta }+T_{\phi }.
\end{equation}%
We note that $T_{\theta }$ and $T_{\phi }$ contribute to both the net normal
and net meridional force density. The force balance
equation that the system must satisfy, in the case of a stable invagination,
becomes%
\begin{equation}
-\mu (T_{\theta }+T_{\phi })\leq \frac{\partial T_{\theta }}{\partial \theta 
}+(T_{\theta }-T_{\phi })\cot \theta \leq \mu (T_{\theta }+T_{\phi }).
\label{FOB}
\end{equation}%

Finally, it is interesting to consider the net force exerted by the sphere
on the entire membrane. The horizontal component of the net force
is zero due to the cylindrical symmetry, and the vertical component is  
\begin{equation}
F_{v}^{mem}=2\pi R^2\int_{0}^{\theta _{max}}(F_{N}\cos \theta -F_{f}\sin \theta
)\sin \theta d\theta .  \label{eq_Fz}
\end{equation}%
If the total vertical force on the membrane is positive, the membrane will
accelerate away from the sphere, and no stable invagination is realized. If
the total vertical force on the membrane is negative, the center or mass of
the membrane will accelerate towards the south pole. Therefore, the stable
invagination is attained only if the total vertical force $F_{v}^{mem}$is
zero. 

\subsection{Hyperelastic strain energies and forces}

The components $T_{\theta }$ and $T_{\phi }$ are not independent, but
are related through stretches in the membrane. Details are provided by models
of elasticity. In order to gain a thorough understanding of the deformation
associated with the invagination, we consider three elasticity models:
Mooney-Rivlin hyperelastic model in the Hookean limit, the Neo-Hookean
limit, and the Gent model. We use the proper Cauchy stress throughout.

In spherical coordinates, the principal stretches tangential to the sphere
are 
\begin{equation}
\lambda _{\theta }=\frac{\partial s }{\partial r},  \label{eq_lambda_th}
\end{equation}%
and 
\begin{equation}
\lambda _{\phi }=\frac{\sin (s/R) }{r/R},  \label{eq_lambda_phi}
\end{equation}%
where $s$ is the distance to the north pole after the deformation, and $r$ is the distance to the north pole in the underformed state. We assume that the deformation is isochoric; then%
\begin{equation}
\lambda _{z}=\frac{1}{\lambda _{\theta }\lambda _{\phi }}.
\end{equation}%
With the stretches $\lambda _{\theta }$ and $\lambda _{\phi }$ defined, we
can find the corresponding stresses from the various models of elasticity.

The hyperelastic incompressible Mooney-Rivlin strain energy density has the
form \cite{Mooney, Rivlin, Audoly}%
\begin{equation}
W_{R}=C_{1}(I_{1}-3)+C_{2}(I_{2}-3),
\end{equation}%
where%
\begin{equation}
2C_{1}+2C_{2}=G,
\end{equation}%
and $G$ is the shear modulus, and%
\begin{eqnarray}
I_{1} &=&\lambda _{\theta }^{2}+\lambda _{\phi }^{2}+\frac{1}{\lambda
_{\theta }^{2}\lambda _{\phi }^{2}}, \\
I_{2} &=&\lambda _{\theta }^{2}\lambda _{\phi }^{2}+\frac{1}{\lambda
_{\theta }^{2}}+\frac{1}{\lambda _{\phi }^{2}}
\end{eqnarray}%
are the two invariants of the left Cauchy-Green deformation tensor.  If $%
C_{2}=0$, it gives the Neo-Hookean energy density 
\begin{equation}
W_{NH}=\frac{1}{2}G(\lambda _{\theta }^{2}+\lambda _{\phi }^{2}+\frac{1}{%
\lambda _{\theta }^{2}\lambda _{\phi }^{2}}-3).
\end{equation}%
In the limit of small strains, we have the Hookean energy density,%
\begin{equation}
W_{H}=2G((\lambda _{\theta }-1)^{2}+(\lambda _{\phi }-1)^{2}+(\lambda
_{\theta }-1)(\lambda _{\phi }-1)).
\end{equation}%
The Gent elasticity model takes into account the finite
extensibility \cite{Gent}, with 
\begin{equation}
W_{G}=-\frac{GJ_{m}}{2}\ln (1-\frac{I_{1}-3}{J_{m}}),
\end{equation}%
where $I_{m}=J_{m}+3$ represents the maximum value of $I_{1}$ allowed by the
system in the absence of failure.  In the limit as  $I_{m}\rightarrow \infty 
$, the Gent model reduces to the Neo-Hookean model.

The tangential force density components $T_{\theta }$ and $T_{\phi }$, in
terms of the stretches, are 
\begin{equation}
T_{\theta }=\frac{h_{0}}{R}\frac{1}{\lambda _{\phi }}\frac{\partial W}{\partial \lambda _{\theta
}},  \label{1}
\end{equation}%
and%
\begin{equation}
T_{\phi }=\frac{h_{0}}{R}\frac{1}{\lambda _{\theta }}\frac{\partial W}{\partial \lambda _{\theta
}}.  \label{2}
\end{equation}%
Substitution of Eqs.~\eqref{1} and \eqref{2} into Eq.~\eqref{FOB}  gives the
inequality in terms of $s(r)$ whose solution provides the detailed
description of the elastic deformation. We note that the shear modulus plays
no role in the invagination.

\subsection{Numerical algorithm and results}

Our goal is to find the stable deformation of the membrane $s(r)$, for which Eq.~\eqref{FOB} must be satisfied. This
requires a non-standard approach, which we now describe. All the lengths are nondimensionalized by the radius of the sphere $R$, so $s/R=\theta$, and $r/R=\bar{r}$. We first prescribe
the portion of the membrane in contact with the sphere by  $\bar{r}_{\max }$, and
the point on the sphere $\theta _{\max }$, where the membrane loses contact.
In this example, we have set $\bar{r}_{\max }=0.9$, $\theta _{\max }=1.8$, which are
in rough agreement with experiments. We also prescribe the initial
displacement $\theta _{0}(\bar{r})$ of the membrane. In particular, we use an
initial displacement of the membrane which is in the neighborhood of the
final displacement, which has been suggested by experiments. Given a
friction coefficient $\mu $, we implement the following psudeodynamics%
\begin{equation}
\theta _{i}^{n+1}=\theta _{i}^{n}+\Delta \theta (\bar{r}_{i})dt,
\end{equation}%
where $\theta _{i}^{n}$ is the position of membrane at undeformed position $%
\bar{r}_{i}$ at time step $n$, and 
\begin{equation}
\Delta \theta (\bar{r}_{i})=\left\{ 
\begin{tabular}{ll}
$0,$ & if $|F_{m}|\leq \mu F_N,$ \\ 
$F_{m}-\mu F_N,$ & if $F_{m}>\mu F_N,$ \\ 
$F_{m}+\mu F_N,$ & if $F_{m}<-\mu F_N.$%
\end{tabular}%
\right. 
\end{equation}%
That is, the point on the membrane does not move if $F_{m}\,$is less than
the maximum frictional force, and the point moves either away from or
towards the north pole if $F_{m}$ overcomes the maximum frictional force,
and the direction of motion depends on the sign of $F_{m}$. We have used 100
equally spaced points $\bar {r}_{i}$ between $0$ and $\bar{r}_{\max }$. All derivatives
are approximated with a centered difference scheme.

After equilibrium is reached, the total vertical force\ exerted on the
membrane by the sphere can be evaluated by Eq.~\eqref{eq_Fz}.  If $%
F_{v}^{mem}$ is different from zero, we adjust the value of the friction
coefficient $\mu $. Explicitly, we increase $\mu $ if $F_{v}^{mem}>0$, and
decrease $\mu $ otherwise, until the total vertical force on the membrane
vanishes.

We remark that the equilibrium solution is highly sensitive to the initial
condition, so is the friction coefficient which stabilizes the invagination.

\begin{figure}[htb]
\centering
	\rotatebox{90}{\hspace{-0.3cm}Hookean}
    \begin{minipage}{0.32\textwidth}
	\includegraphics[width=\linewidth]{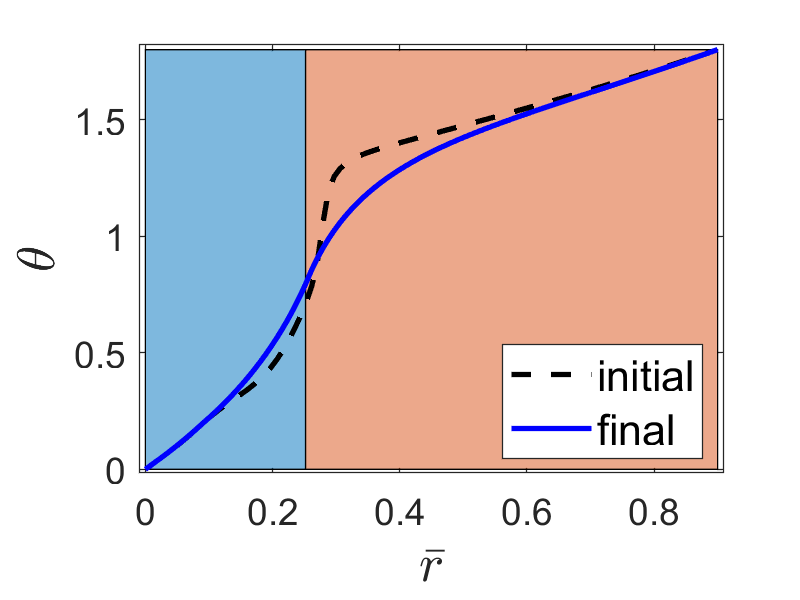} 
   \end{minipage}
    \begin{minipage}{0.32\textwidth}
	\includegraphics[width=\linewidth]{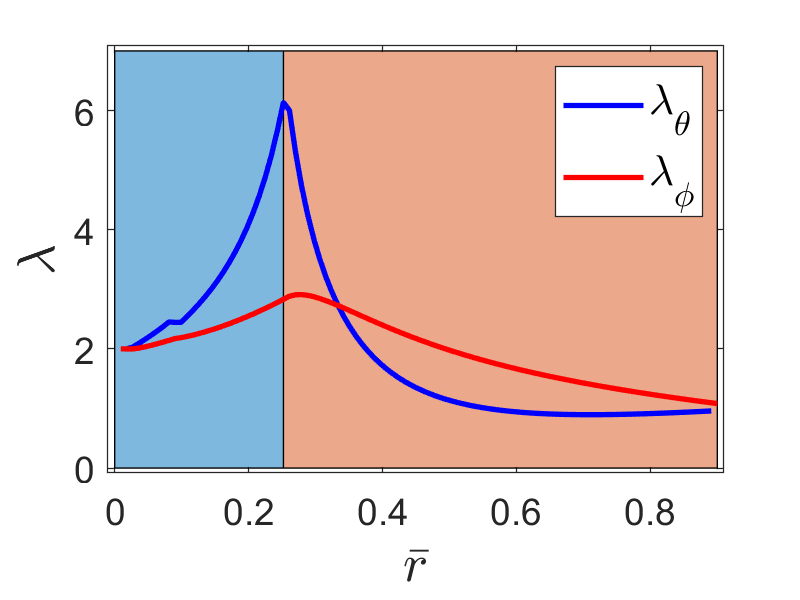} 
	\end{minipage}
    \begin{minipage}{0.32\textwidth}
	\includegraphics[width=\linewidth]{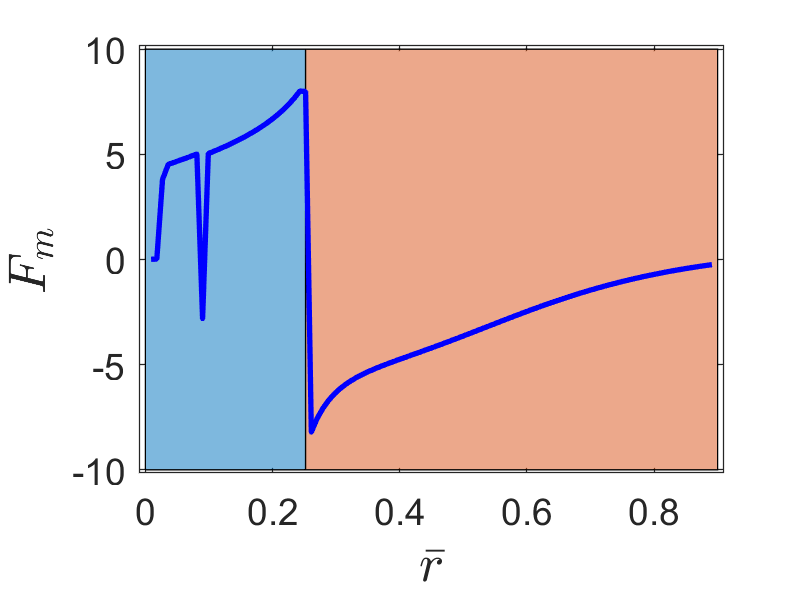} 
	\end{minipage} \newline

\rotatebox{90}{\hspace{-0.8cm}NeoHookean}
\begin{minipage}{0.32\textwidth}
	\includegraphics[width=\linewidth]{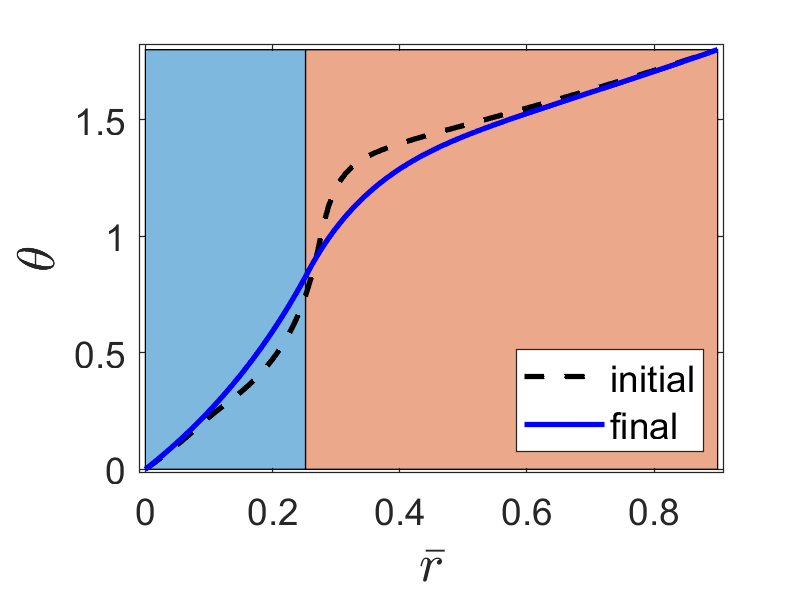} 
\end{minipage}
\begin{minipage}{0.32\textwidth}
	\includegraphics[width=\linewidth]{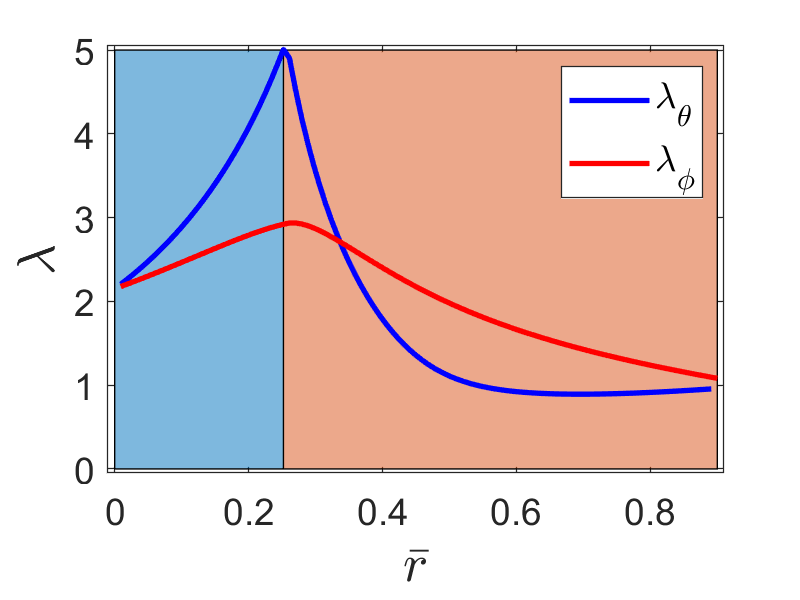}
\end{minipage}
\begin{minipage}{0.32\textwidth}
	\includegraphics[width=\linewidth]{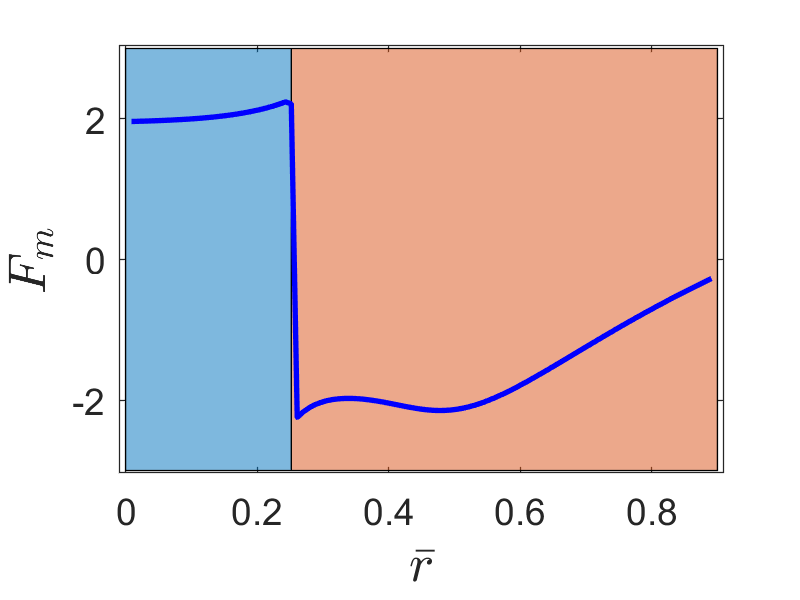} 
\end{minipage} \newline

\rotatebox{90}{Gent}
\begin{minipage}{0.32\textwidth}
	\includegraphics[width=\linewidth]{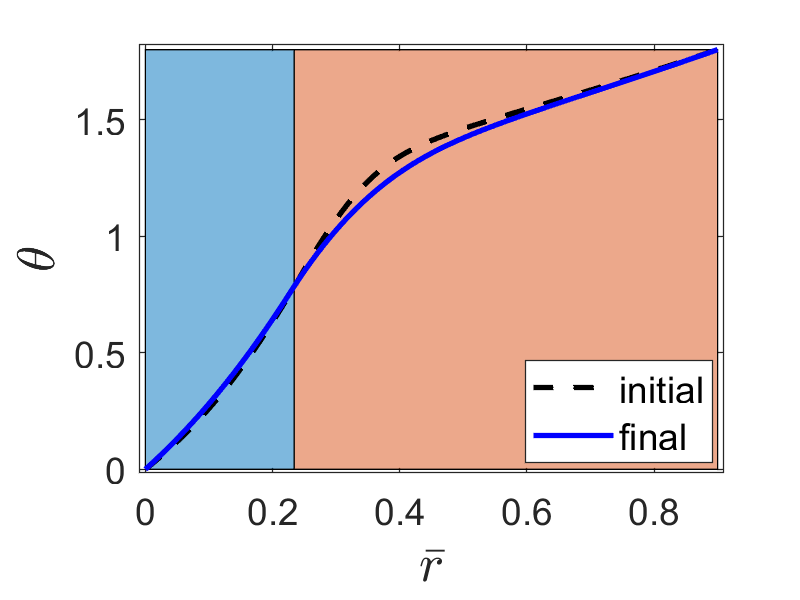} 
\end{minipage}
\begin{minipage}{0.32\textwidth}
	\includegraphics[width=\linewidth]{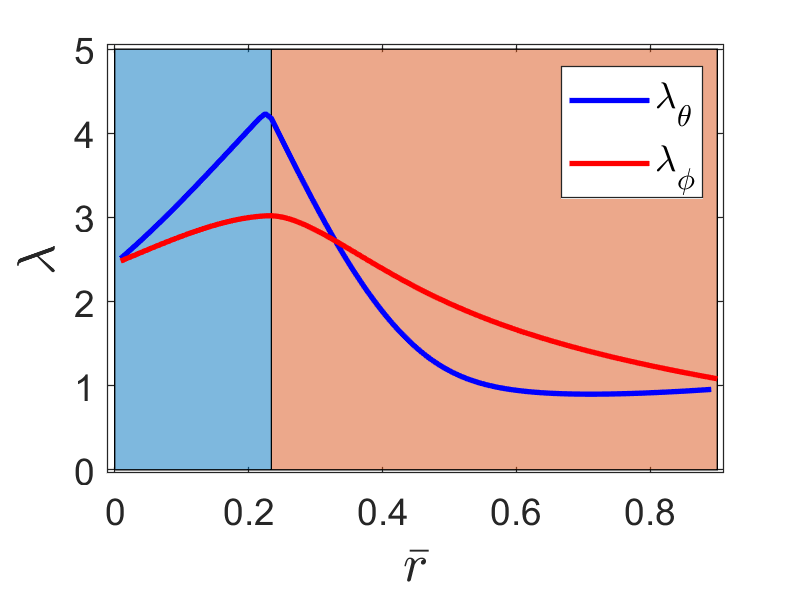} 
\end{minipage}
\begin{minipage}{0.32\textwidth}
	\includegraphics[width=\linewidth]{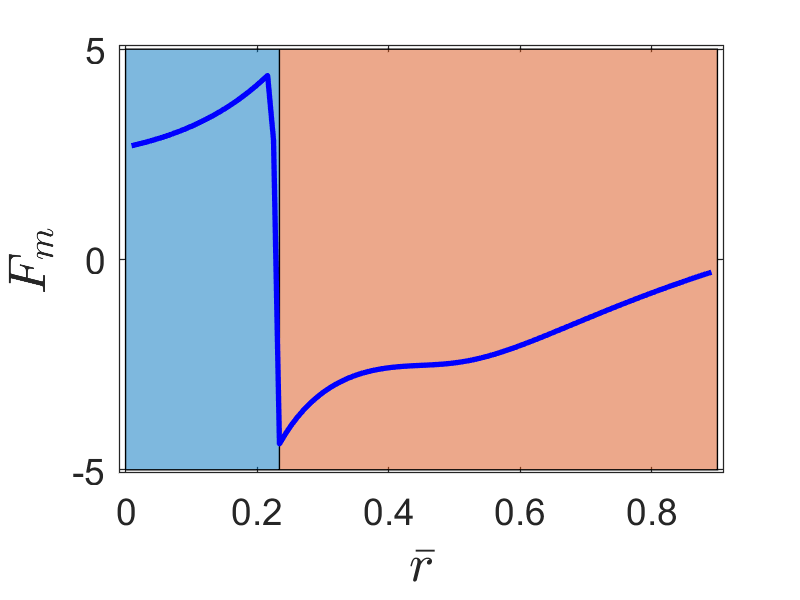} 
\end{minipage} \newline
\caption{Top row: results from Hookean model with $\protect\mu=1.47$.
Middle row: results from neo-Hookean model with $\protect\mu=0.99$. Bottom row: results
from Gent model with $J_m=50$, $\protect\mu=1.1$. Left column: deformation $\protect\theta$ vs.~$\bar{r}$. Middle column: principal stretches $\protect\lambda_{\protect\theta}$ and $ \protect\lambda_{\protect\phi}$
vs.~$\bar{r}$. Right column: elastic force density along meridional direction $F_m$ vs.~$\bar{r}$.}
\label{fig:numerics}
\end{figure}

Fig.~\ref{fig:numerics} shows the numerical results from the Hookean,
Neo-Hookean and Gent models. The results from three different models are
qualitatively the same, with the main difference that they require different
values of friction coefficient to stabilize the invagination. Among the
three models, the Hookean requires the largest friction coefficient, and
Neo-Hookean requires the smallest friction coefficient. The boundary of the
two background colors on the figures marks the change of direction of the
tangential force from elasticity, or equivalently the change of direction of
the frictional force. The boundary also coincides with the location where
the curve $\theta (\bar{r})$ describing the deformation curve changes its
concavity (left column of Fig.~\ref{fig:numerics}), where the two stretches reach their maxima(middle column of Fig.~\ref{fig:numerics}). We note that it is
not at the north pole where the membrane is maximally stretched, but rather
at an angle $\theta $ close to $\pi /4$, where the membrane pulls the
neighboring regions both sides towards each other. The sphere pushes the
membrane upward with increasing force from the north pole to the point with
maximum stretch, where the force changes direction and pulls the membrane
downward towards the point where the membrane loses contact with the sphere, as shown in right column of Fig.~\ref{fig:numerics}.
Remarkably, in the equilibrium configuration, the tangential force is in
balance with the maximum friction force everywhere, except at a finite number
of discrete points where the stretches reach a local extremum and the
friction force changes direction. Any decrease of the friction
coefficient $\mu $ from its thus found equilibrium value will render the
invagination unstable.

\section{The situation in 2D}

It is interesting to ask whether such an invagination can
be realized with an elastic membrane and a circular cylinder. A quick
experiment with a membrane on a circular cylinder shows that the answer
is no. If the cylinder's axis is horizontal, the membrane can't stay on the cylinder under the midplane without an external force, which
is in stark contrast with the spherical case, where the hoop stress from
azimuthal direction can keep the membrane against the sphere below the
equator. Second, in cylindrical geometry, the force balance equation reads 
\begin{equation}
-\mu  T_{\theta }\leq \frac{dT_{\theta }}{d\theta }=-F_{f}\leq \mu T_{\theta
},  \label{eq_2D}
\end{equation}%
where $T_{\theta}$ is the force per length.
The solution for $T_{\theta }$ at any angle $\theta $ is bounded by 
\begin{equation}
T_{0}\exp (-\mu \theta )\leq T_{\theta }\leq T_{0}\exp (\mu \theta ),
\label{eq_2D_sol}
\end{equation}%
where $T_{0}$ is the tension at the top of the cylinder where $\theta =0$.
The total vertical force on the membrane from the cylinder can be written as 
\begin{eqnarray}
F_{v}^{mem} &=&R\int_{0}^{\theta _{max}}(T_{\theta }\cos \theta
-F_{f}\sin \theta )d\theta   \notag \\
&=&R\int_{0}^{\theta _{max}}(T_{\theta }\cos \theta +\frac{dT_{\theta }}{%
d\theta }\sin \theta )d\theta   \notag \\
&=&R\int_{0}^{\theta _{max}}d(T_{\theta }\sin \theta) =RT_{\theta }\sin \theta
_{\max }.
\end{eqnarray}%
From Eq.~\eqref{eq_2D_sol}, since $\theta _{\max }\in (0,\pi )$, the force
on the membrane will be always positive. Thus there is no elastically
stabilized invagination in 2D, where the hoop stress from azimuthal
direction is missing.

\section{Discussion and Summary}

By conducting experiments using regulation size table tennis balls and
dental dams, we have realized elastically stabilized spherical invaginations.
We used two approaches to form the invagination; they gave similar
deformations of the elastic membrane. Friction plays a key role in
stabilizing the system.

Using $|F_{f}|\leq
\mu F_{N}$, we obtained the governing inequality for the elastic deformation
of the membrane on the sphere. The forces at play are expressed in terms of
stretches in the membrane. We used three elastic models to
describe the invagination; crude agreement with experiments was found. 
Although the inequality was used to evolve the dynamics, remarkably, the solution obeys
the equality at almost every point.

Both the experimental data and the numerical solutions indicate that the
maximum stretch of the membrane occurs at $\theta \simeq \pi /4$, where the
both frictional force and vertical component of the vertical force on the
membrane change direction. Due to the highly nonlinear nature of the
problem, the equilibrium solution and the required minimum friction
coefficient are strongly dependent on initial conditions, or equivalently,
on the process establishing the invagination.

We note the close similarity of the invagination of the membrane and the table
tennis ball and the invagination of the membrane and a coin. In the latter
case, the membrane was nearly uniformly stretched on top of the coin, and
the stretch slowly decreases to a relaxed state under the coin. The
invagination was stabilized by the friction force between the vertical edge
of the coin and the membrane. Here, the vertical downward force on the
membrane is from the lower part of the sphere, below the latitude where the
membrane is maximally stretched.

When releasing the ball from the membrane, there is no vertical force to overcome, since the the total vertical force on the sphere exerted by
the membrane is zero. However, there is hoop stress must be overcome. The
retracting force must be sufficiently large to deform the membrane and
increase the radius of the bottom opening to the radius of the sphere. The
sphere may also be released if the elastic and frictional properties of the
membrane changed. In addition to the experiments described here, we
have also realized spherical invaginations using other membranes (balloons,
condoms, liquid crystal elastomers) and spherical rigid bodies (marbles,
push pins, ball chains).  

Here we have only focused on the basic mechanism of the invagination and considered only the
portion of the membrane in contact with the sphere.  Modeling the
entire membrane would give a more complete description of the invagination. 

\section*{Acknowledgments}
 This work was supported by the Office of Naval Research through
the MURI on Photomechanical Material Systems (ONR N00014-18-1-2624) and Air Force contract FA8649-20-C-0011 as part of the STTR AF18B-T003 Electronically Dimmable Eye Protection Devices (EDEPD) program.

\end{document}